%% file: prdl-revised-arxiv.tex
\documentclass[prd,twocolumn,aps,nofootinbib]{revtex4-1}
\usepackage{amsmath,amssymb,amsfonts,float,color,graphicx,graphics,latexsym,placeins,epsfig,multirow,bm,physics,bbold,enumitem,array,footmisc}

\usepackage{hyperref,url}

\usepackage{comment}

\usepackage[caption=false]{subfig}

\newcolumntype{P}[1]{>{\centering\arraybackslash}p{#1}}
\usepackage[compatibility=false]{caption}
\usepackage{subcaption}
\usepackage{varioref}
\usepackage[active]{srcltx}

\expandafter\ifx\csname package@font\endcsname\relax\else
\expandafter\expandafter
\expandafter\usepackage
\expandafter\expandafter
\expandafter{\csname package@font\endcsname}%
\fi
\hypersetup{
	colorlinks=true,       
	linkcolor=blue,          
	citecolor=blue,        
}

\usepackage{fancybox}
\labelformat{figure}{Fig.~#1} 

\newcommand{\B}{\mathcal{B}}
\newcommand{\A}{\mathcal{A}}
\newcommand{\E}{\mathcal{E}}

\newcommand{\ba}{\begin{align}}
\newcommand{\ea}{\end{align}}

\def\3nab{\tilde{\nabla}}

\def\be {\begin{equation}}
\def\ee {\end{equation}}
\def\ba {\begin{eqnarray}}
\def\ea {\end{eqnarray}}

\newcommand{\sfr}[2]
{{\textstyle\frac{#1}{#2}}}

\newcommand{\barray}{\begin{array}}
\newcommand{\earray}{\end{array}}

\newcommand{\sss}[1][0.035cm]{\hspace*{#1}}

\newcommand{\bea}{\begin{eqnarray}}
\newcommand{\eea}{\end{eqnarray}}

\usepackage[dvipsnames]{xcolor}

\begin{document}
\title{Signatures of \emph{asymmetry}: Gravitational wave memory and the parity violation}
%
\author{Indranil Chakraborty}
\email{indranil.phy@iitb.ac.in}
\thanks{Equal contribution to this work.}

\author{Susmita Jana}
\email{susmita.jana@apctp.org}
\thanks{Equal contribution to this work.}
\thanks{Current address: Asia Pacific Center for Theoretical Physics, Postech, Pohang 37673, Korea}
\author{S. Shankaranarayanan}
\email{shanki@iitb.ac.in}

\affiliation{Department of Physics,  Indian Institute of Technology Bombay, Mumbai 400076, India}
\begin{abstract}
{Einstein's equivalence principle suggests a deep connection between matter and spacetime, prompting the question: \emph{if matter violates parity, must gravity?} This letter explores the detection of parity violation 
in gravity using gravitational wave (GW) memory. Gravitational parity violation could be observable through GW amplitude birefringence and large-scale structure correlations. With improved sensitivity, next-generation GW detectors offer unprecedented opportunities to probe these effects. We propose that the integrated cosmological memory (ICM) of GWs, amplified over cosmological distances, can enhance faint parity-violating signatures. Specifically, if GWs from astrophysical events have differing polarization amplitudes, as in Chern-Simons gravity, ICM significantly amplifies this disparity. ICM \emph{uniquely and independently} allows us to test fundamental symmetries, constrain gravity parameters, and gain insights into the interplay of particle physics, cosmology and gravity.} 
\end{abstract}
\maketitle

\noindent {\em Introduction\textemdash}  In general relativity (GR), Einstein established that matter shapes spacetime, causing it to curve, and this curvature of spacetime then dictates how matter moves through the cosmos~\cite{einstein1961relativity}. Given this fundamental interplay, the observed parity violation in the matter sector~\cite{Lee:1956qn,Wu:1957my,Garwin:1957hc}, as established by the Standard Model of particle physics~\cite{Kostelecky:2008ts,Bain:2016lad}, necessitates consideration in the context of gravity~\cite{Brill:1957fx,Jackiw:2003pm}.
  If gravity is influenced by parity-violating matter, consistent with Einstein’s observation, then what observables, detectable in next-generation electromagnetic and GW experiments, could reveal such effects~\cite{Lue:1998mq,Bhattacharya:1999na}? In GW astronomy, parity violation could manifest as frequency dependent amplitude birefringence ---  left- and right-handed circularly polarized wave amplitudes have exponential enhancement/damping --- absent in parity-conserving theories~\cite{Alexander:2007,Okounkova:2021}. In the large-scale structure, parity-violating correlations in galaxy distribution or spin alignment could serve as signatures, distinguishable from standard cosmology~\cite{Cahn:2021,Paul:2024,Inomata:2024}. Unambiguous detection requires precise measurements, novel statistics, and a deeper theoretical understanding of parity violation in gravity. Upcoming GW detectors --- Cosmic Explorer (CE), Einstein Telescope (ET), LISA~\cite{Jenks:2023, Califano:2023, Callister:2023} and galaxy surveys (DESI, BOSS)~\cite{Krolewski:2024,Cahn:2021,DESI:2025zgx} --- offer opportunities to probe these effects.

In GR, asymmetrical motion generates GWs. Since GWs exhibit asymmetries, a passing GW creates secondary weaker GWs. Similarly, these secondary GWs do the same thing --- generate weaker tertiary GWs. If this process continues ad infinitum, it can lead to permanent distortion in the amplitude ---  \emph{GW memory}~\cite{1974-Zeldovich-JETP,Braginsky:1986ia, Braginsky:1987kwo, 1991-Christodoulou-PRL,Thorne:1992,Favata:2010zu,2013-Bieri.Garfinkle-PRD, Flanagan:2015, Tolish_Wald_cosmology:2016,Nichols:2017, Mitman:2020, Jenkins:2021, Mukhopadhyay:2021}. While currently undetectable~\cite{Hubner:2021amk, Boersma:2020gxx}, next-generation detectors may detect GW memory~\cite{Yang:2018ceq, Johnson:2018xly, ET:2019dnz, Islo:2019qht, Reitze:2019iox, Grant:2022bla, Sun:2022pvh, Gasparotto:2023fcg, Xu:2024ybt}. 
Interestingly, the current authors have shown that the {integrated GW memory} over cosmological distances carries a distinctive signature of the underlying cosmological model~\cite{Chakraborty:2024ars}. Specifically, we demonstrated that for high-redshift sources, the amplification of GW memory can increase by up to a factor of 100 for sources around $z \sim 10$, placing it firmly within the detection range of next-generation observatories like CE and ET~\cite{Chakraborty:2024ars}. We refer to this as {integrated cosmological memory} (ICM). This is distinct from the frequency dependent amplitude birefringence for the individual GW sources. 

Given this significant amplification of ICM in GR, we pose the following question: \emph{Could such cosmological events, specifically ICM, provide a means to detect parity violation in gravity?} In theories like Chern-Simons gravity~\cite{Jackiw:2003pm,Alexander:2009tp}, which violate parity, the amplitudes and the quasi-normal mode frequencies of the plus and cross-polarized GWs originated from compact binary coalescence(CBC) are predicted to be unequal~\cite{Barack:2018yly,Bhattacharyya:2018hsj,Shankaranarayanan:2019yjx,Srivastava:2021imr,2023-Wagle.Yunes-PRD,Srivastava:2022slt,Callister:2023,Schumacher:2025plq,Alexander:2024-amp}. If this amplitude difference is minuscule for individual events, could the ICM amplify this disparity to a detectable level in next-generation GW detectors?

Previous studies have explored memory effects in dynamical Chern-Simons (dCS) gravity\cite{Hou:2021-PRD, Hou:2021-JCAP}, primarily within asymptotically flat spacetimes and their connection to asymptotic symmetries~\cite{Strominger_lectures:2017}. However, our investigation is distinct: it is set in \emph{cosmological background} that is asymptotically de Sitter. This fundamentally different asymptotic structure renders the formalisms developed in those earlier works inapplicable here. To overcome this limitation, we employ a newly developed framework for \emph{generic spacetime geometries}, initially established for electromagnetism \cite{2023-Jana.Shanki-PRD} and subsequently extended to GR \cite{Chakraborty:2024ars}. This work utilises this new formalism to probe parity-violating asymmetries specifically relevant to cosmological settings.


\noindent {\em dCS gravity\textemdash} To illustrate this, we consider 4-D dCS gravity, which introduces parity-violating term via non-minimal coupling between an axial scalar field ($\rho$) and pseudoscalar Pontryagin density~\cite{Jackiw:2003pm,Alexander:2009tp}. The action is: 
\begin{equation}
\!\!\!\!\! S= \!\!\int \!\! \mathrm{d}^4x \sqrt{-g} \left[\kappa \, R +\frac{\alpha  \rho}{4} \left( ^{*}R R \right) - \frac{1}{2}\nabla_a \rho\, \nabla^a \rho  + \mathcal{L}_{m}\right] \label{eq:action-new}
\end{equation}
where, $\kappa = 1/(16\pi G)$, $\alpha$ is the coupling constant, $^{*}R R$ represents the Pontryagin density which is a product of Riemann tensor with its dual and couples to the scalar field $\rho$, and $\mathcal{L}_{m}$ corresponds to the matter Lagrangian.  In geometric units, $\alpha$ has the dimensions of length-square and $\rho$ is dimensionless.]
Varying the above action w.r.t. the metric $g_{\mu\nu}$ and scalar field $\rho$, respectively, leads to~\cite{Jackiw:2003pm}:
\begin{equation}
\label{eq:dCSEOMs}
G_{ab} =  \frac{1}{2\kappa}(T_{ab}-2{\alpha}\, {C}_{ab}), \quad
\square \rho = -\frac{\alpha}{4} \, {}^{*}R R.
\end{equation}
Here, $G_{ab}$ is the Einstein tensor, $T_{ab}$ is the stress-energy tensor corresponding to $\mathcal{L}_{m}$ and scalar field $\rho$, and  ${C}_{ab}$ is the Cotton tensor defined as~\cite{2014-Delsate.Witek.etal-PRD}: 
\begin{equation}
\label{eq:cab-weyl}
    {C}_{ab}= 2 {v}^c\, \nabla^d \, \, ^{*}W_{d(ab)c} + {v}^{dc}\, \, ^{*}W_{d(ab)c}.
\end{equation}
where, $^{*}W_{d(ab)c}$ is the dual of the Weyl tensor which encodes the free gravitational field and is crucial for describing parity-violating effects. $v^c:=\nabla^c {\rho}$ is the velocity of the scalar fluid, and $v^{dc}:=\nabla^{(d}v^{c)}$. In Friedmann-Lemaître-Robertson-Walker (FLRW) spacetime, the Weyl tensor vanishes, and the velocity of the scalar field , $v^c$ satisfies $\nabla_c \nabla^{c} {\rho} \equiv \nabla_c v^c = 0$~\cite{Alexander:2009tp}.
  
In the above field equation, the Cotton tensor contains parity violation in the gravity sector. Interestingly, in spherically symmetric spacetimes, the Cotton tensor vanishes~\cite{Jackiw:2003pm}, resulting in the field equations \eqref{eq:dCSEOMs} reducing to those of GR with a minimally coupled scalar field $(\rho)$. Hence, Schwarzschild and FLRW space-times are also exact solutions to dCS gravity~\cite{Jackiw:2003pm,Alexander:2009tp}. However, the metric perturbations about these spacetimes in dCS are different compared to their GR counterparts~\cite{Alexander:2004us,2007-Grumiller.Yunes-PRD,2007-Smith.MarcKamionkowski-PRD}. In particular, the Cotton tensor induces an asymmetry between the left- and right-handed GW polarizations~\cite{Delsate:2018ome}. Specifically, one polarization mode is enhanced compared to the other, leading to birefringence~\cite{2006-Alexander-PLB,Alexander:2024-amp} and potentially providing a solution to baryogenesis~\cite{Alexander:2004us}. 

In the context of compact binary mergers, the dCS model predicts significant deviations from the GR waveform, especially, the amplitude of the two polarization modes are \emph{not equal}~\cite{Yunes:2009hc,Konno10.1143/PTP.122.561,Okounkova:2019dfo, Okounkova:2021, Okounkova:2019zjf,Li:2022grj}. Notably, the presence of the Cotton tensor in the dCS field equations breaks the isospectral relationship between the even and odd parity modes of gravitational perturbations~\cite{Bhattacharyya:2018hsj,Shankaranarayanan:2019yjx,Srivastava:2021imr,Srivastava:2022slt,2023-Wagle.Yunes-PRD}. 
We demonstrate that the amplitude difference of the primary GW from such sources will be amplified due to ICM. We also show that, the difference in ICM for the two polarization amplitudes can provide independent constraints on the dCS parameter ($\alpha$) leveraging the detection capabilities of the next generation GW detectors.
\begin{figure}[!h]
\includegraphics[scale=0.25]{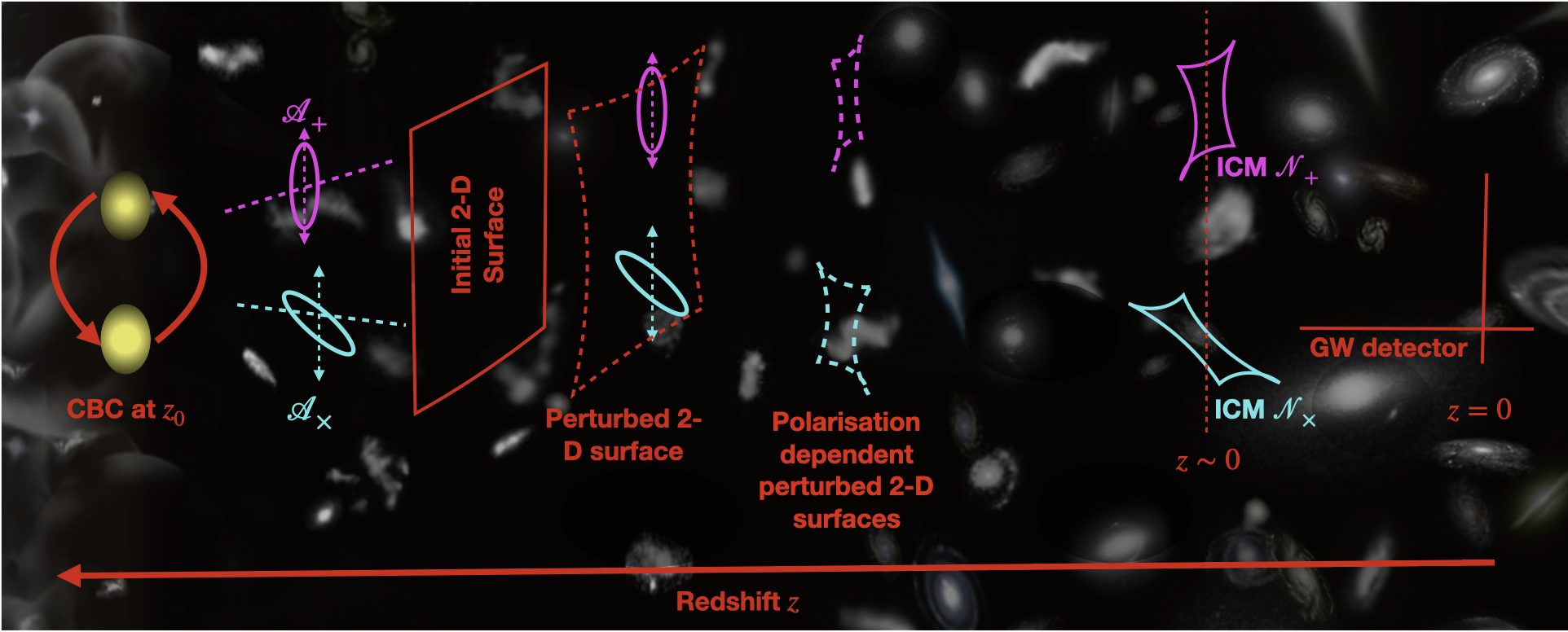}
\caption{Schematic diagram showing how the unequal amplitudes of + and × polarized gravitational waves, a consequence of parity violation in dCS gravity, perturb the background spacetime, resulting in a polarization-dependent amplification of ICM.}
\label{fig:setup}
\end{figure}

\noindent {\em Setup\textemdash} The polarization-dependent amplification of ICM is illustrated  in \ref{fig:setup}. We consider \emph{conformally flat FLRW background}  (with line element $ds^2 =  a^2(\eta) \left(- d\eta^2+ dr^2 + r^2 (d\vartheta^2 + \sin^2\vartheta d\varphi^2)\right)$, where $a(\eta)$ is the scale factor, and $\eta$ is conformal time) which serves as the fundamental metric for the standard $\Lambda$CDM cosmological model. In FLRW space-time, a GW-generating event occurs at $\eta = \eta_0$ (corresponding to $z = z_0$). Due to the parity-violating nature of dCS gravity, the emitted modes of polarized GWs will exhibit different amplitudes \emph{at the source}. As these GWs propagate from $\eta = \eta_0$ to a later time $z = 0$, the primary GWs, with their inherent amplitude differences, generate successive GWs, contributing to the ICM~\cite{Chakraborty:2024ars}. This process is captured on the 2-D surface orthogonal to the direction of propagation of GWs. Our aim is to evaluate the change in the 2-D surface, $N_{ab}$, due to GWs of both the polarization modes as observed by a comoving observer, specifically focusing on how the initial amplitude difference, induced by dCS gravity, evolves and contributes to the ICM.

We evaluate the ICM using comoving observers, which provide a natural reference for studying the large-scale structure and evolution of the universe~\cite{2012-ellis_maartens_maccallum-Book}. To maintain generality and 
\emph{isolate the effects of GWs} on the 2-D surface, $N_{ab}$, we employ \emph{the familiar covariant $1+1+2$ formalism}~\cite{1996-vanElst.Ellis-CQG,1998-Ellis-NATOSci,clarksonlrs}.
In this formalism, the $4-$D spacetime ($g_{ab}$) is split into a timelike direction $u^a$, which is tangent to the comoving observer's worldline in the FLRW background, a spacelike direction $n^a$, and a $2-$D hypersurface ($N_{ab}$), orthogonal to both $u^a$ and $n^a$, i.e.,
\begin{equation}
g_{ab} = h_{ab} - u_{a}u_{b} = N_{ab} + n_{a}n_{b} - u_{a}u_{b}  \, ,
\label{eq:1+1+2}
\end{equation}
In this formalism, any vector or tensor in 3D space ($h_{ab}$) can be expressed as,
\begin{eqnarray}
\mathcal{V}_a &=& N_a\,^c \, n^b \,V_{bc}, \label{eq:vector-2Dprojection} \\
     \mathcal{V}_{ab}  &=& P_{ab}~^{pq}~ \left (h_{(p}~^c ~ h_{q)}~^d - h_{pq}h^{cd}/3\right) {V}_{cd}
  \label{eq:tensor-2Dprojection}
\end{eqnarray}
where $P_{pq}\,^{ab} := (N_p\,^{(a}\, N_q\,^{b)}- \frac{1}{2} N_{pq}\, N^{ab})$. We will denote the derivative of any tensor along $n^a$ \emph{with a hat}. $\delta^a$ is the covariant derivative in 2D-surface orthogonal to $u^a$ and $n^a$.
%
%
 For completeness, details on the 1+1+2 formalism are provided in appendix \eqref{sec:CovariantFormalism}. 

The current authors have recently developed the complete formalism for obtaining ICM in the FLRW background in GR using the 1+1+2 formalism~\cite{Chakraborty:2024ars}. In terms of the 2-D surface $N_{ab}$, ICM is given by~\cite{Chakraborty:2024ars}:
\begin{equation}
\label{def:ICM}
{\cal N}_{ab} = \frac{\Delta N_{ab}}{a^2(\eta)} = 
2 \int_{\eta(z=z_0)}^{\eta(z=0)} d\eta \,\frac{\Sigma_{ab}}{a(\eta)} \, .   
\end{equation}
$\Sigma_{ab}$ is the shear, projected on the $2-$D surface, produced in the background spacetime due to the incoming GWs. $\mathcal{N_{+}}$ and $\mathcal{N_{\times}}$ denote the the plus/cross polarization of the ICM. While these two polarizations exhibit identical behavior in GR, differing only by a phase shift, the dynamical equations governing their evolution in dCS gravity are distinct. This difference arises due to the Cotton tensor, which explicitly encodes the parity-violating nature of the theory \cite{Bhattacharyya:2018hsj,Shankaranarayanan:2019yjx,Srivastava:2021imr,Srivastava:2022slt,2023-Wagle.Yunes-PRD}. Consequently, this leads to a quantifiable difference in the ICM for the two polarizations, which we present in the rest of this letter.


A crucial aspect of this formalism, often overlooked in prior studies of memory effects --- especially those in asymptotically flat spacetimes \cite{Hou:2021-PRD, Hou:2021-JCAP} --- is the role of the ICM. In FLRW background, where both observers and gravitational waves are comoving, the memory signal inherently integrates over its entire propagation history, from the source redshift to detection. As we demonstrate, this results in a \emph{non-trivial redshift dependence} that previous works missed. We highlighted this feature in the earlier work within GR \cite{Chakraborty:2024ars}, demonstrating that the memory's redshift dependence is not simply a straightforward $(1+z)$ scaling.

Another key distinction is our treatment of the \emph{background cosmological evolution}. While the local astrophysical processes generating the memory signal (like its characteristic step-like, DC feature) might be similar across different scenarios, our formalism explicitly tracks how this signal evolves over cosmological time. Because earlier works didn't address this evolution, our framework offers a \emph{novel pathway to utilize memory effects as probes of cosmological models}.



\noindent {\em ICM in dCS theory \textemdash} Given that the equations of motion in dCS gravity \eqref{eq:dCSEOMs} are distinct from GR, we proceed with a series of steps to derive the ICM for this theory.

\noindent {\bf 1. Project tensor quantities on the 2-D surface:} We need to project the dCS gravity equations \eqref{eq:dCSEOMs} onto $N_{ab}$.  While the 2-D projection of $G_{ab}$ and $T_{ab}$ has been derived in Refs.~\cite{2003-Clarkson.Barrett-CQG,Chakraborty:2024ars}, we need to obtain the projection of the Cotton tensor onto the 2-D surface ($\mathcal{C}_{ab}$), as it contains parity violating contributions in the gravity sector. $\mathcal{C}_{ab}$ will be instrumental in determining the parity violating effects on ICM. Substituting Eq.~(\ref{eq:cab-weyl}) into the RHS of Eq.~\eqref{eq:tensor-2Dprojection}, the projected Cotton tensor on 2-D surface is:
\begin{eqnarray}
\label{eq:Cotton-2-space}
\mathcal{C}_{ab}= & \left[N_a\,^{(p}\, N_b\,^{q)} - \frac{1}{2} N_{ab}\, N^{pq}\right] \left[ h_{(p}~^c ~ h_{q)}~^d - \frac{1}{3} h_{pq}h^{cd} \right]  \nonumber \\
& \left[2 {v}^e\, \nabla^f \, \, ^{*}W_{f(cd)e} + {v}^{fe}\, \, ^{*}W_{f(cd)e} \right].
\end{eqnarray} 
From the expression above, we observe that ${\cal C}_{ab}$ is identically zero if the Weyl tensor vanishes, such as in FLRW spacetime. Consequently, 
for such backgrounds, any non-zero contribution to ${\cal C}_{ab}$, and the ensuing polarization-dependent GW memory, arises entirely from the incoming GWs, a direct consequence of the dCS gravity model. 

\noindent {\bf 2. Obtain ${\cal C}_{ab}$:} 
To evaluate $\mathcal{C}_{ab}$, we need to obtain $v^c$.
In FLRW spacetime, Eqs.~\eqref{eq:dCSEOMs} reduce to those of GR, and the equation of motion of $\rho$ behaves as a minimally coupled field. This allows for a gauge choice for $v^a$, specifically, $v^a \propto u^a$ or $v^a = X(\eta) \, u^a$, where $X(\eta)$ is the proportionality factor. The condition $\nabla_{a} \left[X(\eta) u^a\right] = 0$, yields:
\begin{align}
    X(\eta) = X_0 \, a^{-3}(\eta),
    \label{eq:X-theta}  
\end{align}
where $X_0$ is the integration constant. Since ${\cal C}_{ab}$ is identically zero in the unperturbed FLRW background, the first-order perturbations of $v^a$ do not contribute to linear-order perturbations of ${\cal C}_{ab}$. Substituting the above form of $v^a$ in Eq.~\eqref{eq:Cotton-2-space} we obtain $\mathcal{C}_{ab}$.

\noindent {{\bf 3. 2-D projection of Eq.~\eqref{eq:dCSEOMs}}:}~~Substituting the first-order perturbed Einstein tensor and the stress-energy tensor for a perfect fluid and the scalar field, as derived in Ref.~\cite{Chakraborty:2024ars}, into the linearized dCS Eqs.~\eqref{eq:dCSEOMs}, we arrive at the following modified master equation governing the evolution of the 2-D shear tensor $\Sigma_{ab}$:
\begin{equation}
\label{eq:master-eqn-initial}
\ddot{\Sigma}_{ab} - \hat{\hat{\Sigma}}_{ab} +  \frac{\Theta}{2} \dot{\Sigma}_{ab}
+  \left[\dot{\Theta} - (\mu+p) \right] 
\frac{\Sigma_{ab}}{2}  = \epsilon_{c\{a} \, \delta^c\B_{b\}} + \dot{{\Pi}}_{ab}
\end{equation}
where $\B_b$ represents the vectorial component of the magnetic Weyl tensor,
containing information about the primary GW, as defined in Eq.~(\ref{eq:vector-2Dprojection}). $\Pi_{ab} \equiv -2 \alpha {\cal C}_{ab}$ is anisotropic effective stress tensor corresponding to the GW generated by the CBC event in dCS gravity, $\mu$ $(p)$ refer to the energy density (pressure) of the perfect fluid and the scalar field.  For details, please refer to appendix \eqref{B2}.

\noindent {\bf 4. dCS corrected GW memory master equation:}\\
Introducing the retarded null coordinate $u = \eta - r$, the first two terms on the LHS of the above expression~\eqref{eq:master-eqn-initial} cancel~\cite{Chakraborty:2024ars} in the wave zone approximation~\cite{Chakraborty:2024ars}. In the asymptotic limit, we have 
$\partial_u \sim \partial_{\eta}$. Thus, the final form of the dCS-corrected GW memory master equation is:
\begin{equation}\label{eq:master-equation-final}
\dot{\Gamma}_{ab} -\bigg(\frac{\mu+p}{\Theta}\bigg) \Gamma_{ab} = \epsilon_{c\{a} \, \delta^c\B_{b\}} + \bigg(\frac{\mu+p}{\Theta}\bigg) {\Pi}_{ab} \, .
\end{equation}
where, $\Gamma_{ab} := \frac{\Theta}{2} \Sigma_{ab} - {\Pi}_{ab}$. 

The derived master equation, Eq.~\eqref{eq:master-equation-final}, is central to our analysis, and we emphasize the following key aspects: 
First, the parity-violating terms encoded within $\Pi_{ab}$ and $\B_b$ modify the evolution of the 2-D shear tensor $\Sigma_{ab}$. 
Second, like in GR,  Eq.~\eqref{eq:master-equation-final} establishes a relationship between the dCS-corrected 2-D shear tensor $\Gamma_{ab}$, the background spacetime geometry (represented by background quantities $\mu$, $p$, and $\Theta$), and the primary gravitational wave (encoded in $\B_b$ and $Pi_{ab}$). In the absence of the primary GW, when ${\Pi}_{ab} = {\cal B}_a = 0$, the background spacetime remains the unperturbed FLRW at all times, and consequently, $\Gamma_{ab}$ does not evolve. Thus, Eq.~\eqref{eq:master-equation-final} behaves as a forced linear differential equation, with the primary GW acting as the forcing term~\cite{1985-Braginsky-JETP,Siddhant:2020}. 

Third, for the $+$ and $\times$ GW polarization, the plus and cross mode of the above equation reduce to:
\begin{eqnarray}
   \frac{d\, \Gamma_+}{d\eta} + Q(\eta) \Gamma_+ &=& \frac{1}{r} [h_+^{\prime\prime}(\eta) F_1 
   +  h_\times^{\prime\prime}(\eta) F_2 
   ]  \nonumber\\
   \frac{d\, \Gamma_\times}{d\eta} + Q(\eta) \Gamma_\times &=& \frac{1}{r} [h_+^{\prime\prime}(\eta) F_3 
   - h_\times^{\prime\prime}(\eta) F_4 
   ].
   \label{eq:Gamma+x}
\end{eqnarray}
where, prime denotes derivative w.r.t. $\eta$, and
\begin{widetext}
\begin{eqnarray} 
&& Q(\eta) = \bigg(\dfrac{2 {\cal H}^\prime}{3 {\cal H}}-\dfrac{8 {\cal H}}{3}\bigg), \hspace{0.3cm}    
     F_1 \equiv F_1(\eta,\vartheta,\varphi) = \cos(2\varphi) f_1+ \sin(2\varphi) f_2,  \hspace{0.3cm}
      F_2 \equiv F_2(\eta,\vartheta,\varphi) = \sin(2\varphi) f_1 - \cos(2\varphi) f_2,  \nonumber \\
   &&  F_3 \equiv F_3(\eta,\vartheta,\varphi) = \cos(2\varphi) f_3- \sin(2\varphi) f_4,  \hspace{0.1cm}
      F_4 \equiv F_4(\eta,\vartheta,\varphi) = \sin(2\varphi) f_3 + \cos(2\varphi) f_4, \hspace{0.1cm}  \mathcal{H}= \frac{a^\prime}{a}, 
      \hspace{0.1cm}
      f_3=\frac{1}{2}\bigg(1-\frac{\sin^3\vartheta}{2}\bigg), \nonumber \\
   &&   f_1= 1/2 (\cot\vartheta+ \cos\vartheta/2), \hspace{0.15cm} f_2 = \frac{4\alpha H_0}{3a^4} (\cos\vartheta -\cot\vartheta/2) \left(\mathcal{H}-\frac{\mathcal{H}^\prime}{\mathcal{H}}\right), \hspace{0.15cm}
  f_4 = \frac{4\alpha H_0}{3a^4} \bigg(1-\frac{\sin^3\vartheta}{2}\bigg)\, \left(\mathcal{H}-\frac{\mathcal{H}^\prime}{\mathcal{H}}\right).
  \end{eqnarray}
\end{widetext}

Finally, to analyze ICM resulting from CBC events observable by ET/CE, we model the primary GW burst with a sharply peaked amplitude using a Dirac delta distribution: $h_{+/\times}(\eta, \mathbf{n}) = \mathcal{A}_{+/\times}(\mathbf{n}) \, \delta(\eta-\eta_0)$. Here, $\eta_0$ represents the conformal time of the primary GW generation, and $\mathbf{n}$ denotes the direction of propagation. Substituting this ansatz in Eqs.~\eqref{eq:Gamma+x}, converting to redshift, integrating, substituting into Eq.~\eqref{def:ICM}, utilizing the relationship between $\Gamma_{ab}$ and $\Sigma_{ab}$, and retaining only leading order terms, yields:
%
\begin{subequations}
\begin{eqnarray}
\mathcal{N_+}=& \frac{N_+}{a^2}  = h_\oplus \bigg(F_1(z_0)+\frac{\mathcal{A}_{\times}}{\A_+} F_2(z_0)\bigg) \, E^{2/3}(z_0) (1+z_0) \nonumber \\  & \times \int_0^{z_0} \frac{dz}{(1+z) E^{8/3}(z)} \\
 \mathcal{N_\times} =& \frac{N_\times}{a^2}=   h_\oplus \bigg(F_4(z_0)-\frac{\mathcal{A}_{\times}}{\A_+}F_3(z_0)\bigg) \, E^{2/3}(z_0) (1+z_0) \nonumber \\ & \times \int_0^{z_0} \frac{dz}{(1+z) E^{8/3}(z)}
\end{eqnarray} 
\end{subequations}
where, ${\cal N}_{+/\times}$ corresponds to ICM for the $+/\times$ polarization mode, $h_\oplus=\A_p  \, f(\vartheta,\varphi)(1+z)/(3 \, D_L\,H_0^2) $ is the GW amplitude at the earth,  $D_L$ is the luminosity distance [$r=D_L/(1+z)$]~\cite{LIGOScientific:2016wyt}, $H_0$ is the Hubble constant, $E(z)$ is the dimensionless Hubble parameter~\cite{Liu:2024-H0,Croton:2013-H0}. For $\Lambda$CDM, $E(z)=  [\Omega_m (1+z)^{3}+(1-\Omega_m)]$ and $\Omega_m$ is the matter density parameter~\cite{Kamionkowski:2022pkx}.

A significant distinction of our work is the redshift dependence of dCS gravitational wave memory within a cosmological background, contrasting sharply with previous results in asymptotically flat spacetimes. Moreover, our findings in dCS gravity diverge from the GR prediction presented in \cite{Chakraborty:2024ars}, a discrepancy stemming from the inherent parity-violating nature of dCS theory. Crucially, while works like \cite{Hou:2021-PRD, Hou:2021-JCAP} focused on theoretical frameworks without direct gravitational wave signal predictions, our analysis explicitly calculates the differential polarization amplitudes. This concrete prediction highlights the potential of the ICM as a novel observable for probing parity violation with next-generation detectors.


The derived expressions explicitly demonstrate that the amplification of the GW memory is \emph{distinct} for the two polarizations. Notably, even under the hypothetical scenario where the GW source amplitude observed at Earth, the intervening cosmological background evolution, and the overall integrated memory contribution from the source redshift to the present epoch are identical for both polarizations, the initial term within the brackets on the RHS of the expressions --- which directly encapsulates the first-order dCS correction --- exhibits a clear difference between the two polarization states. Consequently, despite the cumulative influence of cosmological propagation being present for both polarizations, the resulting ICM is inherently different for each.

To quantitatively assess the disparity in the ICM between the two polarizations, we introduce a dimensionless parameter $\Delta$, defined by the relation between the source frame amplitudes:
$$\mathcal{A}_{\times} = \mathcal{A}_{+}(1 + \Delta) \, .$$
For dCS gravity, $\Delta \sim \alpha^2/L^4$~\cite{Bhattacharyya:2018hsj}, where $L$ is the characteristic length scale (Schwarzschild radius) of the binary BHs. Utilizing the most recent constraint on $\alpha$ from NICER, $\alpha < (8.5 \, \text{km})^2$~\cite{Perkins:2021}, we estimate $\Delta \sim 10^{-3}$ for binary BHs with component masses around $16 \, M_\odot$, and potentially reaching $0.1$ for smaller BH masses.

\begin{figure}[!hbt]
\includegraphics[scale=0.25]{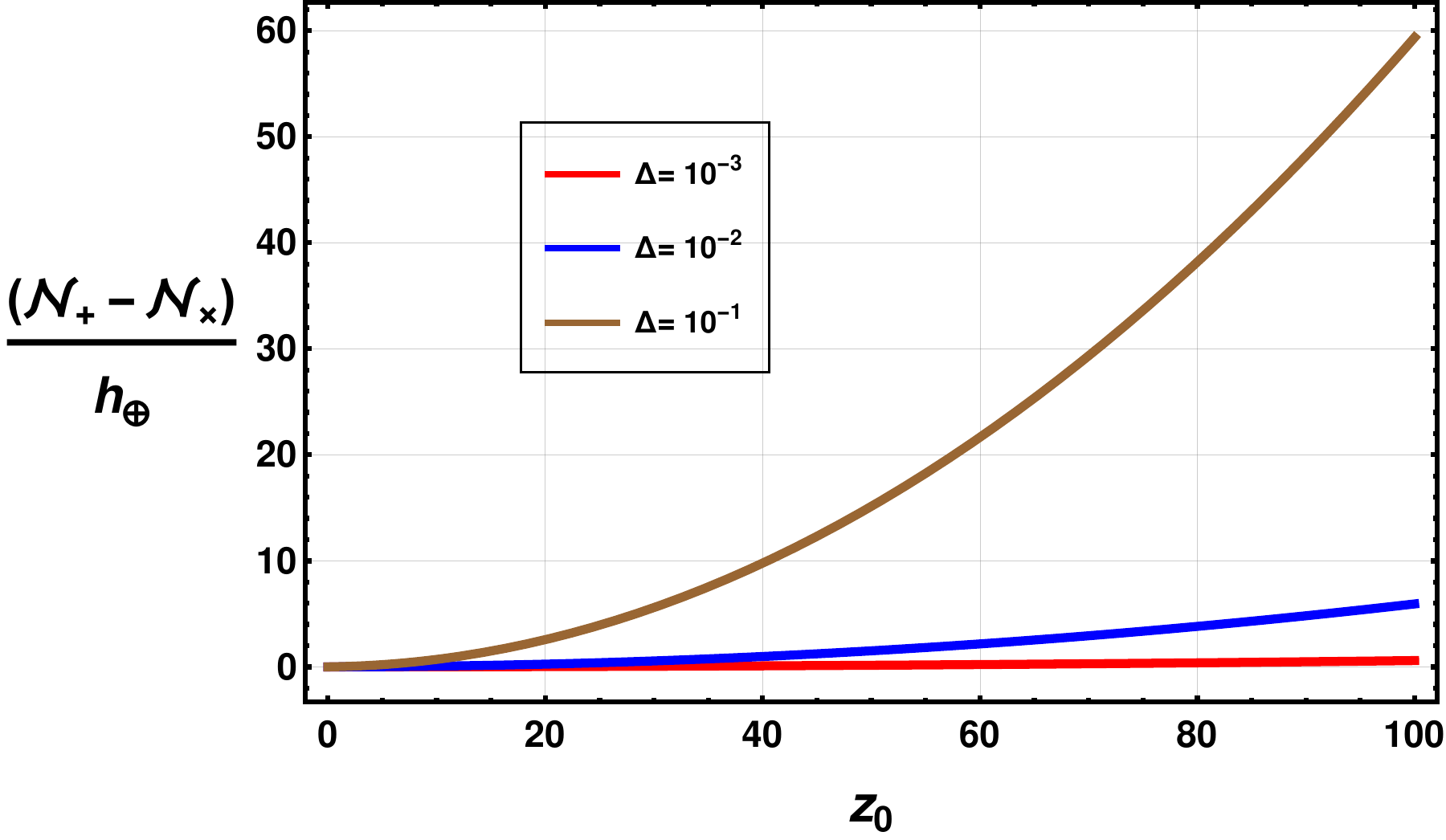}\hfill
\caption{Differential ICM ($\mathcal{N}_+ - \mathcal{N}_\times$) versus source redshift ($z_0$) in dCS gravity for $\Lambda$CDM cosmology, with parameters $\vartheta = \pi/4$, $\varphi = \pi/8$, and $\Omega_m = 0.3$.}
\label{fig:N+diffNx}
\end{figure}

To elucidate its consequence, \ref{fig:N+diffNx} illustrates the difference in the ICM between the plus and cross polarizations as a function of source redshift $z_0$, for a fixed azimuthal angle $\varphi$. While the figure reveals a negligible difference for $\Delta = 10^{-3}$, a noticable and significant disparity emerges even at $\Delta = 10^{-2}$. This demonstrates the potential of next-generation GW detectors to impose stringent constraints on the parity-violating coupling parameter $\alpha$. The behavior of the ICM for individual polarizations ($+$ and $\times$) is detailed in appendix \eqref{sec:appendix-C}.
These results highlight the observable distinctions in the ICM between the two polarizations that will be accessible to next-generation GW detectors, directly implying the following: The measured difference in ICM between the plus and cross polarizations offers a unique and direct probe of parity violation within the gravitational sector, a phenomenon absent in standard GR. This polarization-dependent effect, a direct consequence of the parity-violating parameter $\alpha$ inherent to dCS gravity, enables the establishment of stringent constraints on this parameter and the exploration of the fundamental nature of gravity across cosmological scales. The inherent sensitivity of next-generation gravitational wave detectors, like CE and ET to even small difference in ICM, as shown in our results, opens up new avenues for testing fundamental physics and investigating the cosmological implications of parity violation.

\noindent {\em Cosmological implications\textemdash}The next decade heralds a transformative era for GW cosmology, presenting a unique and independent avenue for exploring the universe. As demonstrated in this work, GW memory stands out as a particularly potent observable, especially in the context of parity-violating theories. While the analysis till now has been restricted to $\Lambda$CDM, it is possible to extend the analysis for generic cosmological model~\cite{Liu:2008-MNRAS,Bargiacchi:2021-QSO-BAO:CPL}): 
\begin{align}    
E(z)=   [\Omega_m (1+z)^3+(1-\Omega_m) f(z)]~,~
{\rm  where,~~} \nonumber \\
f(z)=\exp \left[\int_0^z\, dz'\, \frac{1+w(z')}{(1+z')} \right] \, .
\end{align}
While $\Lambda$CDM assumes $w  = -1$ for the dark energy, however, $w(z)$ can be an arbitrary function~\cite{Kamionkowski:2022pkx}. 
Given that the latest DESI results suggest evolving dark energy~\cite{DESI:2025zgx}, we consider three different  --- CPL~\cite{Chevallier:2000-CPL1,Linder:2002-CPL2}, JBP~\cite{Jassal:2004-JBP} and P2~\cite{Sahni:2002-P2} --- models, with the following equation of state:
{
\begin{eqnarray}
& & w_{CPL}(z)=w_0+\frac{w_1 \, z}{(1+z)};~ 
w_{JBP}(z)=w_2+ \frac{w_3 \, z}{(1+z)^2};~
\nonumber \\ 
& & w_{P2}(z)=-1+\frac{(1+z)[\Omega_1+2\Omega_2(1+z)]}{3[\Omega_2 z^2+(\Omega_1+2\Omega_2)z+0.7]}   
\end{eqnarray}
}
where $w_0, w_1, w_2, w_3, \Omega_1$ and $\Omega_2$ are constants. 
\begin{figure}[!hbt]
\hspace*{-1em}\includegraphics[scale=0.25]{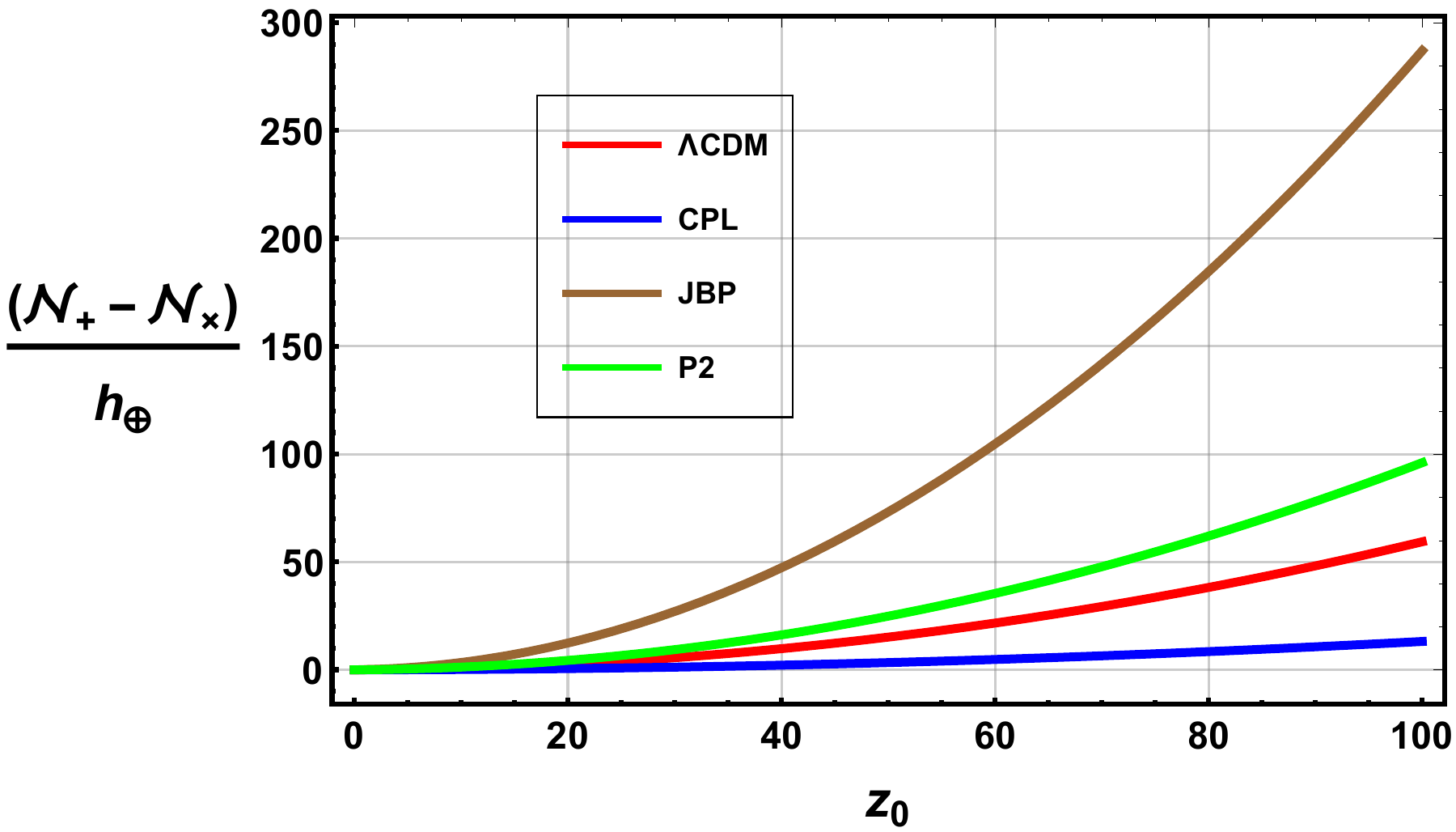}
\caption{Differential ICM ($\mathcal{N}_+ - \mathcal{N}_\times$) versus source redshift ($z_0$) for four cosmological models: CPL, JBP, P2, and $\Lambda$CDM. The model-specific parameters are $\Omega_1=-4.162, \, \Omega_2=1.674$,  $w_0=-1.323,\, w_1=0.745$, $w_2=-1.872,\, w_3=6.628$, $\Omega_m=0.3$, while common parameters are $\Delta = 0.1$, $\vartheta = \pi/4$, $\varphi = \pi/8$. All cosmological parameter values are taken from \cite{Bargiacchi:2021-QSO-BAO:CPL,Liu:2008-MNRAS}.}
\label{fig:n+Nx-cosmo}
\end{figure}
%
\ref{fig:n+Nx-cosmo} contains the plot of the difference in ICM between the two polarizations as function of source redshift $z_0$ for the four cosmological models. 
From the figure we see that the degree of this parity-violating enhancement is sensitive to the underlying cosmological model. Specifically, we see that the differential amplification of ICM, a direct consequence of parity violation, can be significantly enhanced—potentially by a factor of 250 for high-redshift sources depending on the background cosmology. The results clearly show that ICM, modulated by parity-violating effects, offers a distinctive probe into the nature of cosmic acceleration. This approach offers new insights into the universe's expansion and could illuminate fundamental questions about parity violation in gravity and its role in cosmological evolution. 

While parity violation was not a known phenomenon in Einstein's time, Einstein's principle presents a compelling challenge to our understanding of gravity's interplay with \emph{this asymmetry}. As this letter demonstrates, the ICM offers a powerful and independent tool to test parity violation within the gravitational sector. Our analysis has shown that the differential amplification of ICM, a unique signature of parity-violating gravity, can be significantly enhanced over cosmological distances, making it a potentially observable effect for next-generation gravitational wave detectors. Future research will focus on refining these predictions through numerical relativity simulations, exploring the effects of different cosmological models, and investigating potential degeneracies with other astrophysical effects. This comprehensive approach will further solidify ICM as a crucial and independent observable in the search for parity violation in the universe, providing new insights into the fundamental nature of gravity. As fanciful as this proposal seems, we suspect that GW memory detection will become a \emph{unique tool} of cosmology and gravity in the next decade.

\vspace{0.2in}
\noindent {\bf Acknowledgments} The authors thank S. Mahesh Chandran, J. P. Johnson, A. Kushwaha, and S. Mandal for comments on the earlier draft. The work is supported by the SERB Core Research grant (SERB/CRG/2022/002348).
\vspace{0.2in}

\newpage

\begin{appendix}

\section{Semitetrad covariant formalism}
\label{sec:CovariantFormalism} 

This section is a brief revision of the semi-tetrad formalism~\cite{Covariant,2003-Clarkson.Barrett-CQG}, which is used for calculating Integrated Cosmological Memory (ICM) in dynamical Chern-Simons (dCS) theory of gravity. The formalism is developed from the perspective of a real or fictitious observer moving through spacetime along a timelike congruence. The observer's four-velocity is used to decompose the four-dimensional spacetime into a three-dimensional spatial hypersurface and a one-dimensional timelike direction. In spacetimes that possess certain symmetries—such as the Locally Rotationally Symmetric (LRS) class, which includes FLRW models—a preferred spacelike direction naturally arises. This, in turn, allows for a further decomposition of the 3-space into a two-dimensional hypersurface orthogonal to this preferred direction, resulting in a 1+1+2 split of spacetime. All relevant geometric and dynamical quantities, along with the corresponding field equations, are reformulated in terms of this 1+1+2 spacetime decomposition.


Below, we provide an brief overview of this formalism.

\subsection{Semitetrad 1+3 formalism}\label{A1}
Covariant formalism, first proposed in Refs.~\cite{1955-Heckmann-zap, 1955-A.K.Raychaudhuri-PRL}, later were extensively used in relativistic cosmology~\cite{2012-ellis_maartens_maccallum-Book}. In this formalism the $4-$d spacetime is deconstructed w.r.t a fictitious co-moving observer, moving with velocity $u^a = dx^a/d\tau$($\tau$ is the affine parameter), satisfying $u_a u^a = -1$. The spacetime comprises a timelike congruence $\gamma$ and a $3-$d space orthogonal to $u^a$. The $3-$d space is described by the projection tensor $h_{ab}$ that follows:
\begin{equation}
 g_{ab} = - u_{a}\sss u_{b}+ h_{ab},
\end{equation}
iff the 3-space has no twist or vorticity, $h_{ab}$ becomes metric of the $3-$space. The covariant time derivative along the observers' worldlines, denoted by `${\sss\sss^{\cdot}\sss\sss}$' is defined using the vector ${u^{a}}$, as
\begin{equation}\label{dot}
\dot{Z}^{a ... b}{}_{c ... d} = u^{e}\sss\nabla_{e}\sss Z^{a ... b}{}_{c ... d},
\end{equation} 
for any tensor ${Z^{a...b}{}_{c...d}}$. The fully orthogonally projected covariant spatial derivative, denoted by `\sss${D}$\sss' is defined using the spatial projection tensor ${h_{ab}}$, as
\begin{equation}\label{D}
D_{e}\sss Z^{a...b}{}_{c...d} = h^r{}_{e}h^a{}_{f}\sss...\sss h^b{}_{g}\sss h^p{}_{c}\sss...\sss h^q{}_{d}\sss\nabla_{r}\sss Z^{f...g}{}_{p...q},
\end{equation}
The covariant derivative of the 4-velocity vector ${u^{a}}$ is decomposed irreducibly as follows
\begin{eqnarray}
\nabla_{a}\sss u_{b} &=& -u_{a}\sss A_{b} + \frac{1}{3}h_{ab}\sss\Theta + \sigma_{ab} + \epsilon_{abc}\sss \omega^{c},
\end{eqnarray}
where ${A_{b}}$ is the acceleration, ${\Theta}$ is the expansion of ${u_{a}}$, ${\sigma_{ab}}$ is the shear tensor, ${\omega^{a}}$ is the vorticity vector representing rotation and ${\epsilon_{abc}}$ is the effective volume element in the rest space of the comoving observer. The vorticity vector $\omega^q$ is related to vorticity tensor $\omega^{ab}$ as: $\omega^a \equiv (1/2)\, \epsilon^{abc} \, \omega_{bc}$.

Furthermore, the energy-momentum tensor of matter or fields present in the spacetime, decomposed relative to ${u^{a}}$, is given by
\begin{eqnarray} \label{3.Tab}
T_{ab} &=& \mu\sss u_{a}\sss u_{b} + p\sss h_{ab} + q_{a}\sss u_{b} + u_{a}\sss q_{b} + \pi_{ab},
\end{eqnarray}
where ${\mu}$ is the effective energy density, ${p}$ is the isotropic pressure, ${q_{a}}$ is the 3-vector defining the energy-momentum flux and ${\pi_{ab}}$ is the anisotropic stress. The Weyl tensor also is decomposed into electric part $E_{ab}$ and magnetic part $B_{ab}$ as follows,
\begin{eqnarray}
 E_{ab} &=& C_{acbd}\sss u^{c}\sss u^{d} = E_{<ab>}, \label{E}\\
B_{ab} &=& \frac{1}{2} \varepsilon_{ade}\sss C^{de}{}_{bc}\sss u^{c} = B_{<ab>}. \label{H}
\end{eqnarray}
Here, the angle brackets denote orthogonal projections of vectors onto the 3-space as well as the projected, symmetric, and trace-free (PSTF) part of tensors:
\begin{eqnarray} 
\dot{V}_{<a>}& = &h_{a}{}^{b} \sss \dot{V}_{b},\label{angbrac1}\\ 
Z_{<ab>}& =& \bigg({h^{c}{}_{(a}\sss h^{d}{}_{b)} - \frac{1}{3}h_{ab}\sss h^{cd}}\bigg) Z_{cd}.\label{angbrac2}
\end{eqnarray}

\subsection{Semitetrad 1+1+2 formalism}\label{A2}

The $3-$space mentioned in the $1+3$ covariant formalism can be further split into one spacelike direction $e_a$ satisfying $e_a e^a = 1$ and a $2-$d surface orthogonal to both $u^a$ and $e^a$. 
The 1+1+2 covariantly decomposed spacetime is expressed in terms of the projection tensor $N_{ab}$ associated with the $2-$d surface as: 
\begin{equation}\label{2.Nab}
g_{ab} = -u_{a}\sss u_{b} + e_{a}\sss e_{b} + N_{ab},
\end{equation}
where ${N_{ab}}$ ${\left(e^{a}\sss N_{ab} = 0 = u^{a}\sss N_{ab}, N^{a}{}_{a} = 2\right)}$ projects vectors onto the 2-sheets, orthogonal to ${u^{a}}$ and ${e^{a}}$.  We introduce two new derivatives for any tensor ${\phi_{a...b}{}^{c...d}}$:
\begin{eqnarray}
\label{hatderiv}
\hat{\phi}_{a..b}{}^{c..d} &\equiv& e^{f}\sss D_{f}\sss \phi_{a..b}{}^{c..d}, \\
\label{deltaderiv}
\hspace*{-0.6cm}\delta_{f}\phi_{a...b}{}^{c...d} &\equiv& N_{f}{}^{j} N_{a}{}^{l} ... N_{b}{}^{g} N_{h}{}^{c} ...N_{i}{}^{d}  D_{j}\phi_{l...g}{}^{h...i}. \nonumber \\
\end{eqnarray}
Eq.(\ref{hatderiv}) denotes the derivative along the preferred spacelike direction $e^a$ while 
$\delta_f$ in Eq.(\ref{deltaderiv}) gives the 2-space derivative.

The  1+3 geometrical and dynamical quantities and anisotropic fluid variables are split irreducibly as
\begin{eqnarray}
A^{a} &=& \mathcal{A}\sss e^{a} + \mathcal{A}^{a}, \\
\omega^{a} &=& \Omega\sss e^{a} + \Omega^{a}, \\
\sigma_{ab} &=& \Sigma\left(e_{a}\sss e_{b} - \frac{1}{2}\sss N_{ab}\right) + 2\sss\Sigma_{(a}\sss e_{b)} + \Sigma_{ab}, \\
E_{ab} &=& \mathcal{E} \left(e_{a}\sss e_{b} - \frac{1}{2}\sss N_{ab}\right) + 2\sss\mathcal{E}_{(a}\sss e_{b)} + \mathcal{E}_{ab}, \\
B_{ab} &=& \mathcal{H}  \left(e_{a}\sss e_{b} - \frac{1}{2}\sss N_{ab}\right) + 2\sss\mathcal{H}_{(a}\sss e_{b)} + \mathcal{H}_{ab}\,.
\end{eqnarray}
 
The fully projected 3-derivative of ${e^{a}}$ is given by
\begin{eqnarray}
D_{a}\sss e_{b} &=& e_{a}\sss a_{b} + \frac{1}{2}\sss\phi\sss N_{ab} + \xi\sss\varepsilon_{ab} + \zeta_{ab},
\end{eqnarray}
where traveling along ${e^{a}}$, ${a_{a}}$ is the sheet acceleration, ${\phi}$ is the sheet expansion, ${\xi}$ is the vorticity of ${e^{a}}$ (the twisting of the sheet) and ${\zeta_{ab}}$ is the shear of ${e^{a}}$. 

We can immediately see that the Ricci identities and the doubly contracted Bianchi identities, which specify the evolution of the complete system, can now be written as the time evolution, spatial propagation, and spatial constraints of an irreducible set of geometrical variables:
\begin{align}
\label{Dgeom}
\mathcal{D}_{geom}
& = 
 \{\Theta, \sss \mathcal{A}, \sss\Omega, \sss\Sigma, \sss\mathcal{E}, \sss\mathcal{B}, \sss\phi, \sss\xi, \sss\mathcal{A}_{a}, \sss\Omega_{a}, \sss\Sigma_{a}, 
\sss\alpha_{a}, \sss a_{a}, \sss\mathcal{E}_{a}, \sss\mathcal{B}_{a}, \nonumber 
\\ & \sss\Sigma_{ab}, \sss\zeta_{ab}, 
\sss\mathcal{E}_{ab}, \sss\mathcal{B}_{ab}\}
\end{align}

\section{Basic framework in dCS gravity} \label{sec:appendix-B}

\subsection{Actions and field equations} \label{B1}

\noindent The action for dynamical Chern-Simons (dCS) theory is given as,
\begin{eqnarray}
    S= \int d^4x\, \sqrt{-g}\, \bigg(\kappa ~R +\frac{\alpha}{4}~ \rho \,\, ^{*}R R  - \frac{1}{2} [\nabla_
a \rho\, \nabla^a \rho \nonumber \\
+2 V (\rho)] + \mathcal{L}_{m}]\bigg).\label{eq:action-new}
\end{eqnarray}
Now the field equations corresponding to this  becomes, 
\begin{align}
G_{ab} =  \frac{1}{2\kappa}(T_{ab}-2{\alpha}\, {C}_{ab}), \quad
\square \rho = -\frac{\alpha}{4} \, {}^{*}R R.
\end{align}
The Cotton tensor $C_{ab}$ is defined as,
\begin{equation}\label{eq:cab-weyl}
    {C}_{ab}= 2 {v}^c\, \nabla^d \, \, ^{*}W_{d(ab)c} + {v}^{dc}\, \, ^{*}W_{d(ab)c}.
\end{equation}
$v^c:=\nabla^c {\rho}$ is the velocity of the scalar fluid, and $v^{dc}:=\nabla^{(d}v^{c)}$. As shown in the main letter \cite{2025-IC.Jana.Shanki-Manuscript-dCS}, taking the ansatz $v^a=X(\eta) u^a$ in the FLRW background, where $\eta$ is conformal time and $u^a$ is the comoving velocity of the cosmological observer,
    \begin{align}
    X(\eta) = X_0 \, a^{-3}(\eta),
    \label{eq:X-theta}  
\end{align}

In the 2-space $N_{ab}$, the Cotton tesnor becomes,
\begin{eqnarray}
\mathcal{C}_{ab} &= & \left[N_a\,^{(p}\, N_b\,^{q)} - \frac{1}{2} N_{ab}\, N^{pq}\right]\, \left[h_{(p}~^c ~ h_{q)}~^d - \frac{1}{3} h_{pq}h^{cd}\right] C_{ab} \nonumber \\
 &= & \dfrac{5 \bar{X}(t)}{3}~  \bar{\Theta} ~\B_{pq} + \dot{\bar{X}}(t)~ \B_{pq}  - 2 \bar{X}(t) \epsilon_{e < p}  \delta^e \E_{q>}   
\end{eqnarray}

\subsection{Deriving the master equation} \label{B2}

\noindent We treat the tensorial field equation as an effective EFE, one can take the same formalism as introduced in Ref.~\cite{2003-Clarkson.Barrett-CQG}. Thus, the evolution of 2-D shear tensor becomes,
\begin{equation}
    \dot{\Sigma}_{ab} = - \bigg(\frac{2\Theta }{3} + \frac{\Sigma}{2}\bigg) \Sigma_{ab} -\E_{ab} + \frac{1}{2} \tilde{\Pi}_{ab}.
\end{equation}
Here, the $\tilde{\Pi}_{ab} = -2\alpha~ \mathcal{C}_{ab}$. The last term on the RHS arises due to the dCS anisotropic contribution.
Taking a dot-derivative and rearranging some terms, we get
\begin{equation} \label{dot-electric-weyl}
    \dot{\E}_{ab} = - \ddot{\Sigma}_{ab} + \frac{1}{2} \dot{\tilde{\Pi}}_{ab} - \bigg(\frac{2\Theta }{3} + \frac{\Sigma}{2}\bigg) \dot{\Sigma}_{ab} -  \bigg(\frac{2\dot{\Theta} }{3} + \frac{\dot{\Sigma}}{2}\bigg) \Sigma_{ab}.
\end{equation}
From the propagation equation for the 2-D shear tensor we have
\begin{equation} \label{eq:sigma-hat}
    \epsilon_{c\{a} \, \hat{\B}_{b\}}\,^c = -\hat{\hat{\Sigma}}_{ab} + \frac{3}{2}\, \, (\hat{\Sigma} \zeta_{ab} + \Sigma \hat{\zeta}_{ab})
\end{equation}

\noindent The propagation equation for $\zeta_{ab}$ is,
\begin{equation} \label{eq:zeta-hat}
    \hat{\zeta}_{ab} = \bigg( \frac{\Theta}{3} + \Sigma \bigg) \Sigma_{ab} - \frac{1}{2} \tilde{\Pi}_{ab} - \E_{ab}
\end{equation}
Substituting Eq.(\ref{eq:zeta-hat}) in Eq.(\ref{eq:sigma-hat}) gives
\begin{equation} \label{eq:zeta-hat-rearranged}
  \epsilon_{c\{a} \, \hat{\B}_{b\}}\,^c =   -\hat{\hat{\Sigma}}_{ab} + \frac{3}{2}\, \, \hat{\Sigma} \zeta_{ab} +\frac{3}{2} \Sigma \bigg[ \bigg( \frac{\Theta}{3} + \Sigma \bigg) \Sigma_{ab} - \frac{1}{2} \tilde{\Pi}_{ab} - \E_{ab}   \bigg]
\end{equation}
The evolution equation for electric Weyl ${\E}_{ab}$ on $2-$D surface is,
\begin{eqnarray} \label{eq:electric-weyl-evolution}
    \dot{\E}_{ab} -  \epsilon_{c\{a} \, \hat{\B}_{b\}}\,^c +  = & -\frac{1}{2}\dot{\tilde{\Pi}}_{ab}  - \epsilon_{c\{a} \, \delta^c\B_{b\}} - \frac{1}{2} \delta_a \tilde{Q}_b - \frac{1}{2} (\mu+p) \Sigma_{ab}  \nonumber \\ 
   & - \bigg(\Theta+ \frac{3}{2}\Sigma\bigg) \E_{ab} - \bigg(\frac{\Theta}{6}-\frac{\Sigma}{4}\bigg) \tilde{\Pi}_{ab}\, 
\end{eqnarray}
Substituting the first two terms of the LHS of Eq.(\ref{eq:electric-weyl-evolution}) by Eqs.(\ref{dot-electric-weyl}) and (\ref{eq:zeta-hat-rearranged}) respectively, yields
\begin{equation} \label{eq:master-eqn-initial}
    \ddot{\Sigma}_{ab} - \hat{\hat{\Sigma}}_{ab} + \frac{\dot{\Theta}}{2} \Sigma_{ab} + \frac{\Theta}{2} \dot{\Sigma}_{ab} - \frac{1}{2} (\mu+p) \Sigma_{ab} = \epsilon_{c\{a} \, \delta^c\B_{b\}} + \dot{\tilde{\Pi}}_{ab}
\end{equation}

Going on-shell by assuming conformal coordinates in FLRW, one can define a null coordinate by $\tilde{u}= \eta-r$ and assume that at large distances $\partial_\eta\sim -\partial_r\sim \partial_{\tilde{u}} $. Thus, the first and second terms cancel in the RHS of Eq.(\ref{eq:master-eqn-initial}). Defining a quantity $\Gamma_{ab}:=\frac{\Theta}{2} \Sigma_{ab}-\tilde{\Pi}_{ab}$ yields the master equation,
\begin{equation}\label{eq:master-equation-final}
    \dot{\Gamma}_{ab}-\bigg(\frac{\mu+p}{\Theta}\bigg) \Gamma_{ab} = \epsilon_{c\{a} \, \delta^c\B_{b\}} + \bigg(\frac{\mu+p}{\Theta}\bigg) \tilde{\Pi}_{ab}
\end{equation}

\section{ICM profiles for plus and cross polarization} \label{sec:appendix-C}

\begin{figure}[h]
\includegraphics[width=0.7\linewidth]{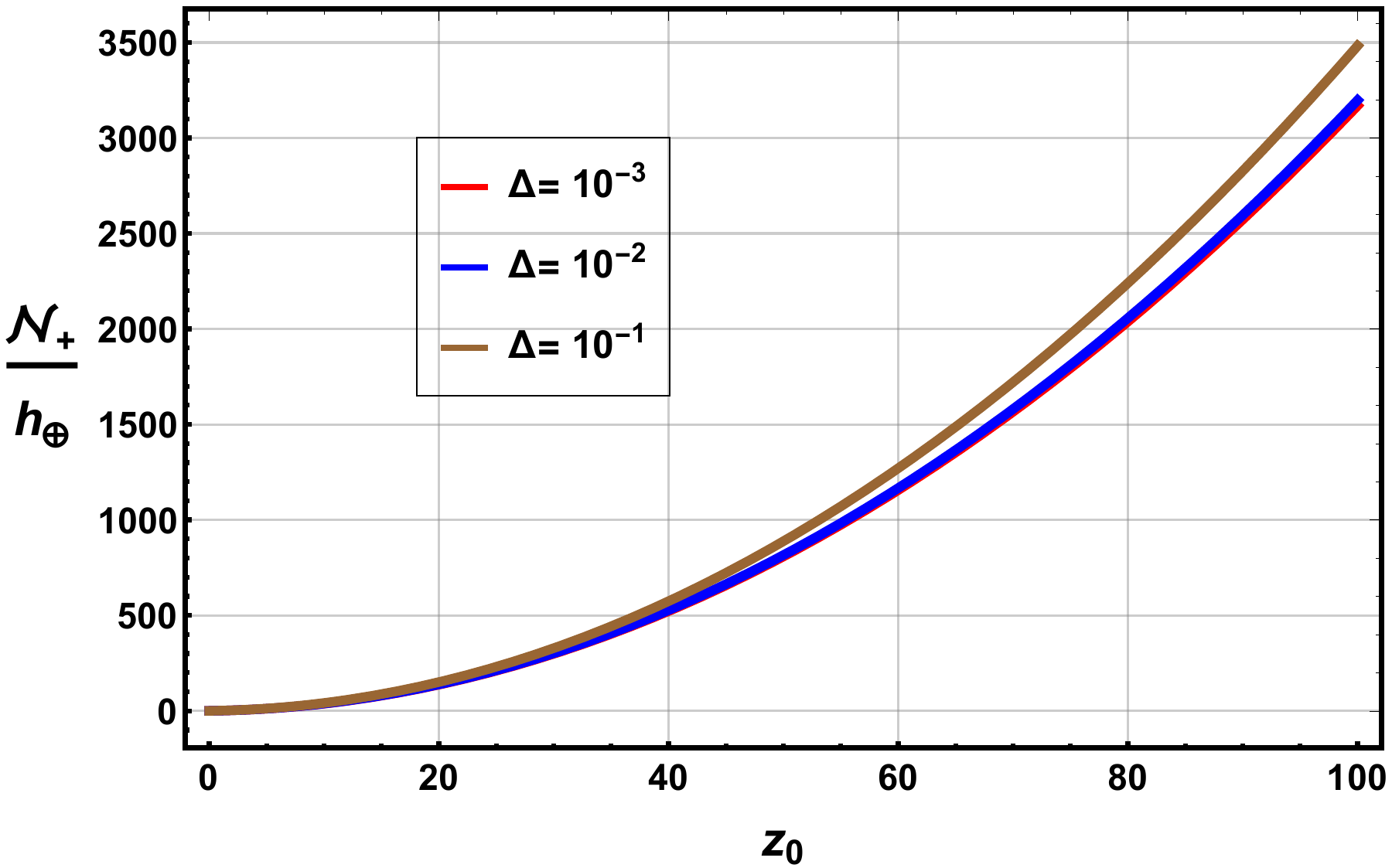} 
\includegraphics[width=0.7\linewidth]{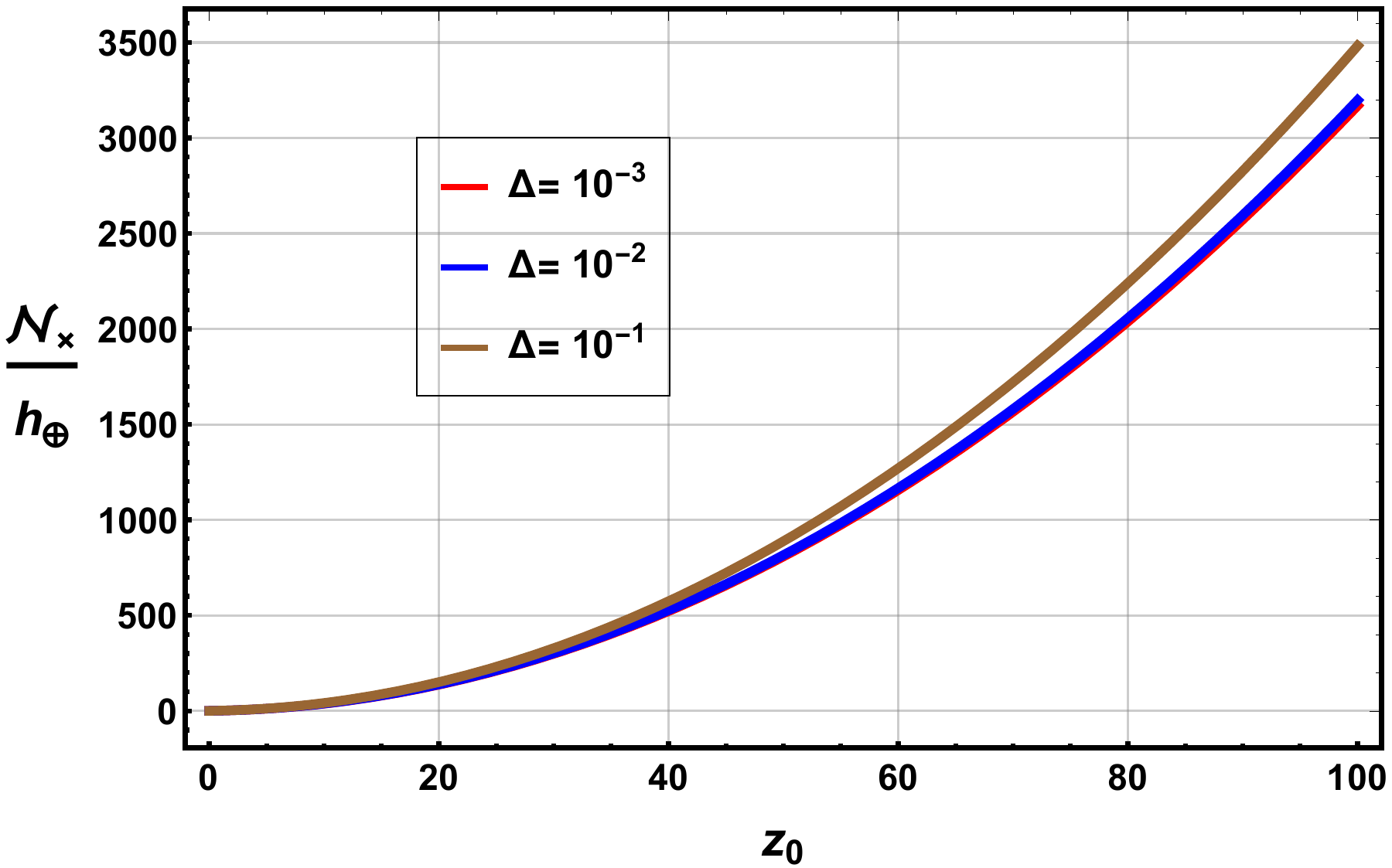}
\caption{ICM for different values of $\Delta$ for $+$ (top) and $\times$ (bottom) polarization modes in dCS gravity for $\Lambda$CDM cosmology. We have set $\Omega_m = 0.3$. In the top (bottom) plot, the value of $\varphi= \pi/4 \,\, (\pi/2)$.}
\label{fig:N+Nx}
\end{figure}

The difference in the two polarizations due to ICM has been demonstrated in the main letter \cite{2025-IC.Jana.Shanki-Manuscript-dCS}.  In \ref{fig:N+Nx}, we plot ${\cal N}_+/h_\oplus$ and ${\cal N}_\times/h_\oplus$ as a function of source redshift $z_0$ for three different values of $\Delta$. From the figure, we observe that 
for all values of $\Delta$, the ICM signal increases monotonically with source redshift. This is consistent with the understanding that, while individual GWs weaken with redshift, the cumulative contributions of successive GWs over cosmological timescales lead to an overall growth in the GW memory signal~\cite{Boybeyi:2024aax,Chakraborty:2024ars}.

\end{appendix}

\input{prdl-revised-arxiv.bbl}
\end{document}

%% file: prdl-revised-arxiv.bbl
%